# An ASM-based Characterization of Starvation-free Systems

Alessandro Bianchi

*Department of Informatics, University of Bari, Bari, Italy*

Via Orabona, 4, 70126 Bari Italy

Tel + 39 080 544 2283

alessandro.bianchi@uniba.it

Corresponding author

Sebastiano Pizzutilo

*Department of Informatics, University of Bari, Bari, Italy*

Via Orabona, 4, 70126 Bari Italy

Tel + 39 080 544 3282

sebastiano.pizzutilo@uniba.it

Gennaro Vessio

*Department of Informatics, University of Bari, Bari, Italy*

Via Orabona, 4, 70126 Bari Italy

Tel + 39 080 544 2283

gennaro.vessio@uniba.it

# An ASM-based Characterization of Starvation-free Systems

Abstract State Machines (ASMs) have been successfully applied for modeling critical and complex systems in a wide range of application domains. However, unlike other well-known formalisms, e.g. Petri nets, ASMs lack inherent, domain-independent characterizations of computationally important properties. Here, we provide an ASM-based characterization of the *starvation-free* property. The classic, informal notion of starvation, usually provided in literature, is analyzed and expressed as a necessary condition in terms of ASMs. Thus, we enrich the ASM framework with the notion of *vulnerable rule* as a practical tool for analyzing starvation issues in an *operational* fashion.

## 1      Introduction

Several formalisms are successfully applied to the development of critical and complex systems in a wide range of application domains, and to their *ex-ante* and *ex-post* analysis aimed at verifying and validating functionality and quality issues. Representing the system-under-study at a high level of abstraction allows developers to focus on algorithmic aspects, rather than on specific realizations of solutions at lower levels. Moreover, the mathematical foundation of formal methods provides complete and unambiguous investigations about the behavior and the properties the system-under-study is required to exhibit.

Some formalisms provide inherent characterizations of properties — in the sense that they can be viewed as independent from the application domain — so that the formal verification of the computationally interesting properties of the modeled systems can be easily conducted. For example, in the Petri net framework [16], a marking $M_i$ is *reachable* from an initial marking $M_0$ if a sequence of transitions transforms $M_0$ into $M_i$. If a marking is not reachable, then the transitions it drives are useless and can be deleted. However, several other formalisms do not provide such features.

Our long-term research is aimed at providing an analogous framework for capturing computationally interesting properties with Abstract State Machines (ASMs) [28]. The goal is to enrich the general body-of-knowledge of the ASM framework and reinforce it as a conceptual tool that developers can find useful and practical in order to analyze system properties in an *operational* fashion. In this paper, "operational" means that the formal specification describes procedurally the system behavior by providing an abstract machine, which can be transformed in an executable form: this is the case of ASM-based models. This notation is usually mentioned in opposition to "declarative" specifications which, instead, state the desired properties by applying a purely descriptive language: this is the case, for example, of temporal logics [37].

With respect to other approaches to the problem of analyzing properties, we focus on ASMs because of the advantages they provide under several viewpoints. When expressivity is considered, ASMs represent a general model of computation which subsumes all other classic computational models [28]. In fact, [13] emphasizes the naturalness with which other computational models, such as Turing machines, can be directly defined as ASM instances without any extraneous encoding (the vice versa is not always true). Secondly, concerning understandability, the ASM approach provides a way to describe algorithmic issues in a simple abstract pseudo-code, which can be translated into a high level programming language source code in a quite simple manner [12]. Thirdly, considering methodological issues, the ASM formalism has been successfully applied for the design and analysis of critical and complex systems in several domains, and a specific development method came to prominence in the last years [13]. Finally, considering the implementation point of view, the capability of translating formal specifications into executable code, in order to conduct simulations of the models, is provided by tools like AsmL [30], CoreASM [22] and ASMETA [25].

This paper specifically deals with the need to specify *starvation-free* systems, here intended as systems comprising processes each capable to make "progress infinitely often" [2]. In order to characterize this feature in terms of ASMs, in [8] and [9] this issue was investigated by introducing the notions of *predicate abstraction over ASM states* and *vulnerable rules*, respectively. The former is here used as a support for the newly defined concept of *risky predicate*; the latter is here more precisely defined. In addition to these novelties, the present paper links all these concepts to the classic, semantic notion of starvation, leading to the proof of a theorem which expresses when an ASM models a starvation-free system. To this end, the informal definitions of starvation, usually proposed in the literature, are here reformulated in form of a necessary condition formally expressed in terms of ASMs. It is worth noting that the elaboration provided in the following is enriched with the application to the well-known Dining Philosophers problem [19] as running example. Moreover, a more realistic scenario, derived from the Mobile Ad-hoc NETwork (MANET) domain, is taken into account. The analysis of this case moves from the case studies presented in [6], [7] and [9], but is elaborated in more details in order to show the applicability of the newly introduced concepts.

The rest of this paper is structured as follows. Section 2 takes into account the literature related to our study. Section 3 deals with the semantic notion of starvation: starting from the discussions provided by some influential works, the common issues are recognized and framed within a necessary condition. Section 4 depicts background knowledge on the ASM formalism. Section 5 is about the notion of starvation from the ASM point of view. Section 6 applies the proposed approach to a MANET case study. Section 7 concludes the paper.

## 2  Related Work

The capability of Abstract State Machines to subsume all other classic computational models has been stated in several works. In [28] Gurevich introduces ASMs, referred to there as *evolving algebras*, as "versatile machines which would be able to simulate arbitrary algorithms"; in [40] Reisig provides an insight into the basic principles of ASMs in order to show their expressive power; in [18] Dershowitz explicitly states that ASMs represent a "general model of computation". Thanks to this generality, an ASM sequential thesis has been proved in [29]: it states that ASMs suffice to capture the behavior of wide classes of sequential systems at any desired level of abstraction. Research efforts have put into extending this thesis to parallel machines [11] and concurrent computations [27], [14]. The latter results seem to comprise a large class of distributed algorithms. Although no theoretical result proves that ASMs suffice to capture the behavior of all classes of distributed algorithms, they have shown to be sufficiently expressive to model concurrency in many applications: network consensus, master-slave agreement, leader election, phase synchronization, load balance, mobile ad-hoc networks, and so on. These observations justify the suitability of the ASM framework in analyzing system properties in a wide range of domains.

In the state of the art, ASMs support both manual and automatic formal verification of systems. Concerning manual analysis, numerous proofs are provided to illustrate how a modeler can verify properties of a given ASM in [13]. Indeed, ASMs are machines equipped with a notion of *run* that lend themselves to traditional mathematical reasoning or manual simulation. These proofs range from simple to complex and are conceived for being used by human experts. Another approach is provided in [24], where a verification calculus based on the Hoare logic is proposed. However, the calculus only considers *partial correctness*, i.e. the result of the computation is what was expected, and is only tailored for a specific class of ASMs.

Moreover, it is worth noting that a logical framework for reasoning about ASMs has been proposed in [43]. However, this framework does not provide operational characterizations of computationally interesting properties. The ASM notion of run is very helpful for supporting the practitioners' work, independently from the possibility of developing automatic verification mechanisms. Nevertheless, since it requires human effort, the manual approach does not offer absolute guarantee and is error-prone.

Concerning automatic analysis, several examples of model checking techniques applied to ASMs exist. In [17] the authors introduce a way to translate ASM specifications into the SMV (Symbolic Model Verifier) language. The goal is to link the workbench they built to the model checker SMV. Another application of model checking techniques to ASMs, based on the CoreASM modeling framework, has been proposed in [23]. In particular, CoreASM-based specifications are transformed into models, written by using the Promela modeling language, that can be verified with the model checker Spin. A tool, AsmetaSMV, that enriches the ASMETA framework with the capabilities of the model checker NuSMV to verify properties of given ASM models is presented in [4]. Moreover, in [39] the authors present an approach to verify ASM models, specified in terms of the Asmeta language, using the model checker Bogor. However, all these solutions present the drawback due to the Turing-completeness of the ASM formalism [29]: properties are, in general, undecidable, so the formal verification of ASM specifications cannot be fully automatized [42]. In fact, an algorithm capable of verifying a specific configuration of a given ASM would be able to verify that a certain configuration, e.g. a halting one, is reachable by a Turing machine expressed by means of an ASM. Since the halting problem for Turing machines is undecidable, such an algorithm cannot exist. For this reason, the translation of the given ASM into the input required by the adopted model checker may cause a loss of expressive power.

Our long-term study is aimed at proposing an approach to the property analysis, entirely within the ASM framework, so that the previous limitations can be overcome. On one hand, we want to provide operational characterizations of properties, so that the manual analysis can be perceived more practical when reasoning about the systems' behavior. On the other hand, since the translation of the ASM under study into a less expressive model is not needed, these properties can be investigated whilst preserving the expressiveness of the model before the application of usual model checking techniques.

For what specifically concerns the general discussion about starvation, our research starts from some of the most important contributions in the literature: the three seminal papers by Dijkstra [20], Lamport [33] and Alpern and Schneider [2], as well as the three recent works by Tanenbaum [44], Pnueli, Podelski and Rybalchenko [38] and Baier and Katoen [5]. These six contributions are used in the next Section as a baseline for a precise definition of starvation.

Although these six works are not recent, the problem of starvation is still actual. Nowadays, the need to address this issue is stringent in several domains. In Cloud systems, see for example [31] and [34], virtualization is used to provide a large variety of resources to end-users, through Internet. Unfortunately, since the number of resources is limited, submitted jobs may experience very long waiting time or may never be executed, so resulting in a starvation problem. A second example concerns the security domain, where several distributed applications can suffer from the so-called Denial of Service (DoS) attacks. According to [36] and [21], a DoS attack causes starvation because it deprives the resources of a target victim to provide services to legitimate users. Research in this domain is typically conducted from an empirical point of view, however we believe that, due to their complexity, such issues can benefit from a formal approach.

Note that, in this paper, we capitalize two concepts recently introduced: *predicate abstraction over ASM states* and *vulnerable rule*. The concept of predicate abstraction (defined in [8]) is here used without modification to characterize the predicates yielding the risk of starvation. On the other hand, the original definition of vulnerable rule ([9]) is not sufficient for our purposes because it derives from the analysis of a particular case; therefore, the notion is here reformulated in order to encompass generality.

Finally, it is worth mentioning that the running example used in the rest of the paper, i.e. the Dining Philosophers problem, is elaborated with respect to both its general statement [19] and its discussion in terms of ASMs [13]. The Dining Philosophers problem is very effective for discussing starvation. Nevertheless, in the present paper, the general discussion is enriched with the application of the proposed approach to a real case in the MANET domain. The case study moves from [9] and redefines the model so that starvation can be detected and solved in accordance with our framework.

## 3   The Notion of Starvation

The literature does not provide a univocal, formal definition of starvation: different authors discuss starvation from different perspectives, in general in an informal manner. At high description level, starvation is described as an accident occurring in multi-process, concurrent systems which hampers a process to continue its proper computation. A process starves because it requires the access to an external resource which is never available. In some cases, for example in communication (sub-)systems, the required resource is represented by a message a process waits for.

### 3.1 A General Necessary Condition for Starvation

In order to capture the classic, semantic notion of starvation, we take into account some classic authors, as well as some more recent views. Dijkstra explains starvation using the Dining Philosophers metaphor, so referring it to the literal case of a person who is dying from hunger: "*although all individual eating actions take only a finite period of time, a person may be kept hungry for the rest of his days*" [20]. Other authors, e.g. Lampert, and Alpern and Schneider, focus on the need to express starvation by means of its negation: a multi-process system is starvation-free when each *"process eventually receives service"* [33] or when each *"process makes progress infinitely often"* [2], respectively. Within the context of operating systems, Tanenbaum states that: "*some policy is needed to make a decision about who gets which resource when. This policy, although seemingly reasonable, may lead to some processes never getting service even though they are not deadlocked*" [44], so expressing starvation in opposition to deadlock. Pnueli, Podelski and Rybalchenko frame starvation-free within the discussion of fairness properties: "*the* [starvation-free] *property relies on justice assumptions that none of the processes idles forever in some location*" [38]. Finally, Baier and Katoen, when treating model checking techniques, observe that: "*each waiting process will eventually enter its critical section*" [5].

Although syntactically different, the six views above share several common semantic issues. They are abstracted in the following:

**Necessary Condition (general statement).** A starvation situation could arise in a system with multiple processes, say $p_1, p_2, \ldots, p_n$, if both the following sub-conditions hold simultaneously:

g.1   The execution of a process $p_i$ ($i = 1, 2, \ldots, n$) requires a service provided by (at least) a process $p_j$ ($j = 1, 2, \ldots, n, j \neq i$).

g.2   The process $p_i$ is forced to wait for the desired service.

The condition above is rather straightforward; however, it is tailored to our purposes in that it encompasses the informal notion of starvation provided by the previous literature. Note that starvation only emerges when multiple processes interact with each other. Therefore, shifting the focus on sequential programs, instead of multi-process systems, implies that starvation cannot occur *a priori*.

Sub-condition (g.1) states the dependency of the process $p_i$ upon the execution of $p_j$; sub-condition (g.2) states the inactivity of $p_i$ while waiting for $p_j$ completion, i.e. $p_i$ is idle in some state. Only the logical conjunction of (g.1) and (g.2) represents a necessary condition. In fact, in the case the computation of each process does not strictly depend on other processes, no blocking event can occur. Moreover, if process $p_i$ is not forced to wait in an idle state, its computation can evolve, so starvation does not occur.

The condition is not sufficient because, even in the case that the two sub-conditions hold simultaneously, a starvation situation would not necessarily occur. For example, we can imagine a process waiting for a resource for a time $t$. The user could suspect starvation is occurring, but no one can be sure that in a time $T > t$ the process will obtain the desired resource. In other words, the starvation-free property is semi-decidable. In order to avoid starvation, one of the two sub-conditions (g.1) or (g.2) must be negated.

One more remark concerns the services blocking the execution of the process, which are intended in a broad sense. They can be the result of a computation that produces, for example, data consumed by the waiting process. They can also be messages for signaling events, availability of resources, changes of state, and so on.

Finally, note that sub-conditions (g.1) and (g.2) are expressed referring to *one* process $p_i$ and *one* process $p_j$, but this case can be easily generalized to sets of processes.

## *3.2     Running Example: Dining Philosophers*

The Dining Philosophers problem, due to Dijkstra [19], is one of the most illustrative examples in the field of concurrency for explaining starvation (and deadlock). Five philosophers are sitting around a table with a bowl of spaghetti in the middle. For the philosophers, life consists only of two moments: thinking and eating, rigorously with two forks. More precisely, since each philosopher has a pair of a right fork and a left fork, (s)he behaves as follows: (s)he thinks till both forks become available, eats for a certain amount of time, then stops eating (putting back both forks on the table) and starts thinking again. The problem is that in between two neighboring philosophers there is only one fork: each philosopher shares his/her right and left fork with his/her corresponding right and left neighbors, respectively. Therefore, at any time, only two of five philosophers can eat. Even if metaphoric, this scenario is very significant: philosophers can be regarded as processes, forks as shared resources, and thinking and eating as computing activities.

The scenario above is affected by the risk of starvation: if no-preemption holds, i.e. the resource consumption cannot be interrupted, or the actions of all philosophers are synchronized in such a way that for (at least) a philosopher both forks can be not available at the same time, then a starvation situation could arise. In fact, this situation satisfies the general necessary condition introduced above. The set of philosophers represents a multi-process system; the execution of the eating process of each philosopher requires the services provided by the others (namely, the forks released by the neighbors); each philosopher is forced to wait for these services.

For the sake of completeness, it is worth noting that if resource holding holds, i.e. the processes continue to have the acquired resources while waiting for the other requested resources, then the same scenario is affected by the risk of deadlock: each philosopher picks up his/her right fork and waits for the left fork to become available. In this way, all philosophers indefinitely wait for each other to release the possessed fork.

## 4     The Abstract State Machine Framework

In the following, background about the ASM formalism is provided together with the notion of predicate over ASM states introduced in [8]. The ASM-based Dining Philosophers problem is then expressed accordingly.

### *4.1     Background*

Abstract State Machines are finite sets of so-called *rules*, which transform *abstract* states. The concept of abstract state extends the usual notion of *state* occurring in finite state machines: it is an algebra over some signature, i.e. a domain of objects of arbitrary complexity with functions and relations defined on them. On the other hand, the concept of rule reflects the notion of *transition* occurring in traditional transition systems, and it is extended to also allow parametrized rule calls $R(a_1, a_2, \ldots, a_n)$ with actual parameters $(a_1, a_2, \ldots, a_n)$ coming with a rule definition of the form $R(a_1, a_2, \ldots, a_n) = body$, where

*body* is a rule. When the arity of a rule is 0, it assumes the form **if** *condition* **then** *updates*.

A complete presentation of the ASM framework can be found in [13]: the book includes much of the previous literature on ASMs, mainly by Gurevich (e.g. [28], [29]). In the following, we only provide some definitions useful for our purposes.

**Definition 1 (Signature).** A *signature* $\Sigma$ is a finite collection of function names. Each function is characterized by an arity which is the number of arguments that function takes. Every signature is assumed to contain the constants *true*, *false* and *undef*.

The concept of function is to be intended in the mathematical sense: according to [13], "one may imagine functions as represented by tables". Partial functions are turned into total functions by using *undef*. Instead, relations are expressed as particular functions that always evaluate to *true*, *false* or *undef*.

**Definition 2 (State).** A *state s* is an algebra over the signature $\Sigma$, i.e. a non-empty set of objects of arbitrary complexity together with interpretations of the functions in $\Sigma$.

**Definition 3 (Location).** A *location* is a pair, $(f, (t_1, \ldots, t_n))$, of a function name *f*, together with values for its arguments $(t_1, \ldots, t_n)$.

Locations abstract the notion of memory unit: the current configuration of locations together with their values determines the current state of the ASM. A location-value pair, *(loc, t)*, then represents the single *update*.

**Definition 4 (ASM).** An *Abstract State Machine M* is a tuple ($\Sigma$, $S$, $R$, $P_M$). $\Sigma$ is a signature. *S* is a set of states. *R* is a finite set of *rules*, basically of the form **if** *condition* **then** *updates*, which transform the states in *S*. In a rule, *condition* is a first-order formula whose interpretation can be *true* or *false*; whereas *updates* is a finite set of assignments of the form $f(t_1, \ldots, t_n) := t$, whose execution consists in changing in parallel the value of the specified functions to the indicated value. $P_M \in R$ is a distinguished rule of arity zero, called the *main rule* or *program* of the machine, which represents the starting point of the computation.

For convenience, the general form of ASM rules encapsulates simple assignments of the form $t := t'$, which are intended as rules of the form **if** *true* **then** $t := t'$. Note that other rule constructors exist; however, their discussion is outside the scope of this paper.

The execution of an ASM consists in iterating computational steps. An *ASM computational step* in a given state consists in executing all rules whose condition evaluate to *true* in that state. Since different updates could affect the same location, it is necessary to impose a consistency requirement. A set of updates is said to be *consistent* if it contains no pair of updates referring to the same location. Therefore, if the updates are consistent, the result of a computational step is a transition of the machine from the current state to another. Otherwise, the computation does not yield a next state. An ASM *run* is so a (possibly infinite) sequence of steps: the computational step is iterated until no more rule is applicable.

In contrast to traditional computational models, in which states are represented by symbols belonging to finite alphabets, the representation of states as configurations of locations makes the ASM computation more complex. In order to better understand the semantics of the ASM states with respect to the computational behavior of the modeled system, it is worth remarking that each ASM state can be characterized by one or more predicates over the states. In [8] the following definition is provided:

**Definition 5 (Predicate over ASM state).** A *predicate $\phi$ over an ASM state s* is a first-order formula defined over the locations determining *s*, such that $s \vDash \phi$.

The state *s* satisfies the predicate $\phi$; in other words, the predicate $\phi$ holds in the configuration of locations which defines *s*. Note that a state can satisfy several predicates and, conversely, a predicate can be satisfied by several states. Two examples of this statement are discussed in [8]. In that work it was observed that an ASM could starve even if the computation evolves through different states, so it is difficult to recognize effective progress. In order to overcome this problem, predicates over ASM states were proposed as an abstraction framework capable of capturing the semantics of these states: each predicate allows one to focus on the subsets of locations that turn out to be interesting for verification purposes. The aim of the concept of predicates over the states is so to allow modelers to assign behavioral meanings to states.

We remark that the notation of parametrized rule supports the mechanism of procedure call: the interpretation of a parameterized rule is to define its semantics as the interpretation of its body with the formal parameters replaced by actual arguments. Note that parameterized rules support the declaration of *local* functions, so that each call of a parameterized rule works with its own instantiation of its local functions.

The aforementioned notions refer to the definition of *basic* ASMs. However, there exist several generalizations, e.g. *parallel* ASMs and *Distributed* ASMs (DASMs), which in turn can behave synchronously (*sync* ASMs) or asynchronously

(*async* ASMs) [13]. Parallel ASMs are basic ASMs enriched with the **forall** construct, for expressing the simultaneous execution of sets of rules satisfying a given condition [26]. For the purposes of the present work, we take into account only sync and async ASMs, that are able to capture the formalization of multiple agents acting in parallel, i.e. using a global system clock, or concurrently, i.e. using different clocks, respectively [13].

**Definition 6 (DASM).** A *Distributed Abstract State Machine D* is a set of pairs ($a$, $ASM(a)$) of pairwise different agents, elements of a possibly dynamic set *Agents*, each executing its own underlying basic ASM $ASM(a)$.

It is worth remarking that the classic definitions for Distributed ASMs refer to "a finite *indexed set* of single-agents" [28] or "a *family* of pairs of pairwise different agents" [13]. However, for the purposes of the present work, such distinction is unessential, so we can only refer to "*set* of pairs".

The single computational step of an individual agent $a$ is called *move*: it is the application of the rules declared in $ASM(a)$ in its current state.

**Definition 7 (Distributed run).** A *distributed run* of a DASM is a partially ordered set ($M$, $\leq$) of moves.

For the sake of completeness, it is worth noting that, according to [28], a distributed run must satisfy three conditions: *finite history*, i.e. each move has only finitely many predecessors; *sequentiality of agents*, i.e. the moves of a single agent are linearly ordered; and the *coherence condition*, which implies that all linearizations of a finite run have the same final state. The partial order of moves determines which agent's move comes before and is only restricted by the consistency condition, which is indispensable.

The relationship between an ASM and its environment (more generally, other ASMs in the case of DASMs) is established by the functions occurring in the ASM and, more precisely, by the class these functions belong to. According to [13], a primary distinction concerns *basic* functions, intended as elementary, and *derived* functions, whose values are defined in terms of other (basic or derived) functions, but neither the ASM nor the environment can update them. In fact, derived functions are automatically updated as a side effect of the updates over the functions from which they derive. In addition, basic functions are classified into *static*, whose values never change during a run, and *dynamic*, for which values change as a consequence of the updates executed by the ASM or by the environment. Furthermore, dynamic functions can be distinguished into other sub-classes. They are called *controlled*, if directly updated only by the ASM. They are *monitored*, if directly updated only by the environment, and only read by the ASM. The functions that are both controlled and monitored are called *shared*. The *out* functions are updated, but never read by the ASM.

Finally, [13] encloses development phases from requirements capture to implementation in a unique ASM-based method. Requirements can be captured by constructing so-called *ground models*, i.e. representations at high level of abstraction that can be graphically depicted. Starting from ground models, hierarchies of intermediate models can be constructed by *stepwise refinements*, leading to executable code: each refinement describes the same system at a finer granularity. The method then supports both verification, through formal proof, and validation, through simulation.

## 4.2 Running Example: ASM-based Dining Philosophers

The ASM-based model of Dining Philosophers here described is based on [13], but with the inclusion of aspects tailored to our purposes, such as the predicates over the states. The scenario can be simply modeled by a DASM $D$ composed by a homogeneous set of agents: each of them behaves according to the same underlying ASM. More precisely, we have a set of *philosophers* = $\{p_1, \ldots, p_5\}$, representing the agents of the system, and a set of *forks* = $\{f_1, \ldots, f_5\}$, representing their shared resources. Each $p_i$ is modeled by the ASM *PhilosopherProgram*($p_i$).

The functions occurring in the signature of each ASM are:

- *rightFork*: *philosophers* → *forks*, which is a static function indicating a philosopher's right fork;
- *leftFork*: *philosophers* → *forks*, which is a static function indicating a philosopher's left fork;
- *owner*: *forks* → *philosophers* ∪ {*undef*}, which is a dynamic shared function denoting the current user of a fork.

The main rule *DiningPhilosophers* of $D$ is:

```
DiningPhilosophers = {
    rightFork(p₁) := f₁
    leftFork(p₁) := f₅
    rightFork(p₂) := f₂
    leftFork(p₂) := f₁
    rightFork(p₃) := f₃
    leftFork(p₃) := f₂
    rightFork(p₄) := f₄
    leftFork(p₄) := f₃
    rightFork(p₅) := f₅
    leftFork(p₅) := f₄

    forall fᵢ in forks do
        owner(fᵢ) := undef
```

>       **forall** $p_i$ **in** *philosophers* **do**
>           *PhilosopherProgram*($p_i$)
> }

The main program *DiningPhilosophers* assigns values to the static locations: each philosopher $p_i$ has fork $f_i$ on his/her right and fork $f_{i-1}$ on his/her left, except for $p_1$ that has fork $f_5$ on his/her left. Then, *DiningPhilosophers* assigns *undef* to each shared location representing the holding of a fork. This means that, at the beginning of the computation, for each $f_i$ in *forks*, *owner*($f_i$) = *undef*, i.e. each philosopher does not hold any fork. Finally, *DiningPhilosophers* runs all the basic ASMs representing each philosopher.

Since the behavior of all philosophers is the same, only one ASM is here described. The basic ASM *PhilosopherProgram*($p_i$), shown below, is composed by two rules, that are RULE 1 and RULE 2:

>       *PhilosopherProgram*($p_i$) = {
>           **RULE 1:**
>           **if** *owner*(*rightFork*(**self**)) = *undef* ∧ *owner*(*leftFork*(**self**)) = *undef* **then** {
>               *owner*(*rightFork*(**self**)) := **self**
>               *owner*(*leftFork*(**self**)) := **self**
>           }
>           **RULE 2:**
>           **if** *owner*(*rightFork*(**self**)) = **self** ∧ *owner*(*leftFork*(**self**)) = **self** **then** {
>               *Eat*(**self**)
>               *owner*(*rightFork*(**self**)) := *undef*
>               *owner*(*leftFork*(**self**)) := *undef*
>           }
>       }

Note that in the ASM above the keyword **self** allows an agent to identify itself within the set of agents. More precisely, an agent *a* interprets **self** as *a*.

The computation of each ASM can evolve through five states:

s.1     (*owner*(*rightFork*(**self**)) = *undef*) ∧ (*owner* (*leftFork*(**self**)) = *undef*);

s.2   $(owner(rightFork(\mathbf{self})) = undef) \wedge (owner(leftFork(\mathbf{self})) = p_{i-1})$;

s.3   $(owner(rightFork(\mathbf{self})) = p_{i+1}) \wedge (owner(leftFork(\mathbf{self})) = undef)$;

s.4   $(owner(rightFork(\mathbf{self})) = p_{i+1}) \wedge (owner(leftFork(\mathbf{self})) = p_{i-1})$;

s.5   $(owner(rightFork(\mathbf{self})) = \mathbf{self}) \wedge (owner(leftFork(\mathbf{self})) = \mathbf{self})$.

Initially, each *PhilosopherProgram*($p_i$) is in state (s.1), i.e. for each philosopher both *owner*(*rightFork*(**self**)) and *owner*(*leftFork*(**self**)) evaluate to *undef*. These states are characterized by the following predicates over the states:

- `thinking`: $\neg(owner(rightFork(\mathbf{self})) = \mathbf{self} \vee owner(leftFork(\mathbf{self})) = \mathbf{self})$. The philosopher is thinking, so (s)he is waiting for both forks to become available. This predicate holds in states from (s.1) to (s.4);
- eating: $owner(rightFork(\mathbf{self})) = \mathbf{self} \wedge owner(leftFork(\mathbf{self})) = \mathbf{self}$. The philosopher is eating, so (s)he has obtained both forks. This predicate holds only in state (s.5).

Note that taking both forks means activating the eating process. In the model above, RULE 2 states that each philosopher, after obtaining both forks, executes the *Eat* rule, then releases them. Since the eating process is outside the resource allocation problem driving the risk of starvation, the *Eat* rule does not need to be further specified.

Ideally, in a fair computation, each agent executes alternately the two rules above to get and later to release the desired forks. In fact, even if ASM rules are executed in parallel by definition, RULE 2 can be performed if and only if RULE 1 has been previously executed.

It is worth remarking that, thanks to the partial order of moves guaranteed by the definition of distributed run, the possible updates of two agents over a same shared location at the same time is not allowed ([13], [14]). Therefore, in the Dining Philosophers DASM, the access to the shared forks is exclusive, so no further scheduling policy is needed.

A final remark concerns the resource holding issue. Note that in our DASM no rule allows a philosopher to grab one fork and wait for the other one, so the resource holding condition does not hold and deadlock can never occur.

## 5     ASMs for Handling Starvation

In order to handle starvation, [38] points out the need to express multi-process systems by means of adequate computational models. Moreover, it acknowledges that starvation lies behind the situation of a state-based model that cyclically returns to the same state. Both observations are taken into account in our view, so that our purpose is to provide an operational characterization of starvation in a state-based fashion.

### 5.1     *ASM-based Starvation-freedom*

We here reformulate the condition provided in Section 3.1 in terms of ASMs.

**Necessary Condition (ASM-based statement).** A starvation situation could arise in a DASM with multiple agents, say $a_1, a_2, \ldots, a_n$, if both the following sub-conditions hold simultaneously:

a.1    A move of an agent $a_i$ ($i = 1, 2, \ldots, n$) requires a service provided by (at least) an agent $a_j$ ($j = 1, 2, \ldots, n, j \neq i$).

a.2    The agent $a_i$ is forced to wait for the desired service.

The transition from the previous general condition to this specific one is purely syntactic: in the following a more precise characterization is provided to better understand it. Firstly, according to the discussion in Section 3.1, starvation is only considered in multi-process systems, where it emerges from the interaction among agents. This is why we exclude from our considerations all basic ASMs representing the behavior of sequential algorithms, or non-interacting with the environment, and only consider DASMs. In fact, although parallel ASMs are able to represent the behavior of parallel algorithms ([11], [26]), they are not prone to starvation: [26] explicitly states that they only deal with single-agent systems; instead, DASMs serve to represent multi-agent systems [13].

Secondly, sub-condition (a.1) states the dependency of the run of ASM($a_i$) on ASM($a_j$); sub-condition (a.2) states that the computation of ASM($a_i$) cannot proceeds as long as it waits, i.e. $a_i$ is idle in some location. In order to operatively treat these issues, it is necessary to express the concepts of *dependency* and *idleness* in terms of ASMs.

Dependency of an agent upon another agent can be expressed in terms of rules and, in particular, in terms of the functions involved in the conditions guarding them. As mentioned above, since starvation emerges in multi-process systems, we need to consider functions expressing the dependencies among agents. With respect to the classification of functions recalled in Section 4.1, static functions surely do not impact starvation: their values never change during a run, so they cannot express such dependencies. Analogously, out functions are not involved in starvation analysis, because their values are produced by the ASM, but never used. Monitored and shared functions surely need special attention, because their presence indicates that the ASM behavior is affected by the other agents. More precisely, we can state that given two ASMs in a DASM — say *ASM($a_i$)* and *ASM($a_j$)*, respectively — if there exists a function *f* such that it is monitored or shared for *ASM($a_i$)* and controlled for *ASM($a_j$)*, and there exists a move *m* in $a_i$ such that *m* includes the application of a rule whose condition is guarded by $f$, then the computation of $ASM(a_i)$ *depends* on the computation of $ASM(a_j)$. Unfortunately, this observation only concerns *direct* dependencies, but does not take into account *indirect* ones. Let $ASM(a_i)$ be a machine, and let $f_c$ and $f_m$ be a controlled and a monitored function, respectively, belonging to $ASM(a_i)$. If the values of $f_c$ are updated as a result of the updates executed by some rule over $f_m$ and $f_c$ is used in some guard, then the satisfiability of those guards is evaluated as $f_m$ is used in them. We say that $f_c$ is *determined* by $f_m$. Conversely, all controlled functions that are not determined by other functions can be excluded from our analysis: their values can be managed by the ASM they belong to, because they are set inside it. Finally, detailed analysis should also be done when derived functions appear. In fact, since these functions cannot be managed inside the ASM, but their values are automatically determined by other functions, the latter must be investigated. We recursively define the functions expressing both direct and indirect dependencies among agents as follows:

**Definition 8 (Risky function).** A *risky function* is: (*i*) a monitored or shared function; or (*ii*) a controlled or derived function determined by a risky function.

A *risky location* is a pair of a risky function name together with values for its arguments. The definition of risky function is then useful for expressing the concept of idleness. In fact, if $ASM(a_i)$ is characterized by (at least) a rule $r$, which is guarded by some risky function, and the computation of the other agents does not allow $r$ to be executed, then the computation of $ASM(a_i)$ cannot evolve. Regarding this aspect, an important issue is related to the granularity used for defining the states of the ASM, in accordance with the level of abstraction realized by the adopted stepwise refinement (see the final remark in Section 4.1). If the states are expressed at a coarser granularity, then the non-execution of *r* forces the ASM to remain in its current state. Instead, if the states are expressed at a finer granularity, then the ASM can go through several intermediate states without carrying out any appreciable computational progress: we call this behavior *cyclical return*. In order to capture the concept of cyclical return we use predicate abstraction. In fact, a predicate over the states whose truth value remains unchanged even when state transitions occur is able to represent the "waiting for something" situation [8] and, again, the dependency of an agent upon the other agents. We can express the dependency of predicates analogously to the dependency of functions through the following:

**Definition 9 (Risky predicate).** Let *D* be a DASM including, among the others, the agents $a_i$ and $a_j$. A *risky predicate* $\phi_r$ over a state *s* of $ASM(a_i)$ is a first-order formula defined over (at least) a risky location in *s*, such that $s \models \phi_r$ and its truth value also depends on $ASM(a_j)$.

The dependency of the truth value of a risky predicate in an agent upon the computation of the other agents states that the condition guarding the rule *r* above, guarded by some risky function, can be satisfied not only as an effect of the computation of the ASM *r* belongs to, but also as an effect of the computation executed by the other ASMs. When the computation of the other agents allows *r* to be executed, then the cyclical return can end, so the risky predicate changes its value and the computation can evolve. This is summarized by the concept of *vulnerable rule* [9]. In the present paper it is more precisely defined as follows:

**Definition 10 (Vulnerable Rule).** Let $\phi_r$ be a risky predicate such that a state $s$, which satisfies $\phi_r$, i.e. $s \models \phi_r$, exists. A rule is said to be *vulnerable* if it is characterized by the following features:

f.1   Its condition includes a logical conjunction with one or more risky functions.

f.2   The truth value of $\phi_r$ only changes as the result of the execution of the updates of some vulnerable rule in $s$.

The presence of risky functions, stated by feature (f.1), represents the interactions between the ASM and the other agents. This implies that the evolution of the computation of at least one agent strictly depends on the other agents. Feature (f.2) formalizes the waiting. More precisely, the presence of a cyclical return to states characterized by the same value for a risky predicate over the states implies that there is a (possible infinite) waiting for a desired external event.

The role of vulnerable rules in driving a risk of starvation is stated by the following:

**Theorem.** *Let* D *be a DASM composed by the set of pairs* $(a_i, ASM(a_i))$. *If each* $ASM(a_i)$ *is without vulnerable rules, then* D *is starvation-free.*

*Proof.* In order to prove the statement above we verify that the absence of vulnerable rules implies that at least one of the two sub-conditions (a.1) or (a.2) of the ASM-based necessary condition for starvation is negated.

If $ASM(a_i)$ is without vulnerable rules, then for every rule in $ASM(a_i)$ at least one of the two features (f.1) or (f.2) of the definition of vulnerable rule does not hold.

If (f.1) does not hold, then all rules in each $ASM(a_i)$ are guarded by conditions without risky functions. Since risky functions are the unique functions implementing the dependency of the agent $a_i$ upon the other agents, their absence states that the moves of $a_i$ are independent from the moves of the other agents. Therefore, the negation of the feature (f.1) negates the sub-condition (a.1).

In order to analyze the second feature, let's note that (f.2) establishes that in state *s*, which satisfies $\phi_r$, only the execution of one or more vulnerable rules can change the truth value of $\phi_r$. Therefore, negating (f.2) implies that in state *s* at least one more rule *r* can be applied and *r* is non-vulnerable. When *r* is executed its update changes the truth value of $\phi_r$, so the computation of $a_i$ can evolve. Therefore, the negation of the feature (f.2) negates the sub-condition (a.2).    □

In conclusion, the lack of vulnerable rules in every ASM means that one of the necessary sub-conditions for starvation does not hold. In particular, at least one of the following characteristics holds for all rules: (*i*) no condition depends on a risky function, so the model is built on ASMs whose computation does not depend on external events; (*ii*) there is (at least) an update that surely changes the truth value of a risky predicate, so the computation can always evolve.

Note that, since starvation-free is semi-decidable, the vice versa is not always true: the possible presence of vulnerable rules in a DASM does not necessarily imply starvation.

## *5.2    Running Example: ASM-based Dining Philosophers Analysis*

The analysis of the model in Section 4.2 clearly shows that it is affected by the risk of starvation: for each *PhilosopherProgram*($p_i$), RULE 1 is a vulnerable rule that can make the philosopher starve because both features (f.1) and (f.2) are satisfied. The rule can be applied only when the current state is (s.1) which satisfies the `thinking` predicate. This predicate is risky because defined over a shared risky location and its truth value depends on locations that can be set by other ASMs. Feature (f.1) holds in that the condition guarding RULE 1 is the logical conjunction of two instances of the shared risky function *owner*; feature (f.2) holds in that there are no non-vulnerable rules changing the truth value of `thinking`.

If both forks never become available, the condition guarding RULE 1 will never be satisfied, so the ASM will cyclically return to one of the states characterized by the `thinking` predicate over the states, i.e. (s.1), (s.2), (s.3), (s.4).

The ASM changes its state every time an update is executed by the neighboring philosophers over the shared locations, i.e. when the neighbors change the value of *owner*(*rightFork*(**self**)) or the value of *owner*(*leftFork*(**self**)). However, the change of state does not necessarily involve the computational progress; in other words, the ASM computation could not evolve towards the state characterized by the `eating` predicate over the states. Indeed, only state (s.1) allows the vulnerable rule to be executed and then reaching the state (s.5).

Conversely, RULE 2 is not vulnerable. It can be applied only when the current state is (s.5). State (s.5) satisfies the `eating` predicate which is not risky because its truth value only depends on the execution of RULE 2 of the same ASM. In conclusion, the satisfaction of its condition is always guaranteed: when the ASM is in state (s.5), the locations *owner*(*rightFork*(**self**)) and *owner*(*leftFork*(**self**)) surely evaluate to **self**, so the computation always evolves towards (s.1).

After the identification of the risk of starvation in RULE 1, it is necessary to overcome it. This can be accomplished by properly re-defining the algorithm by means of an *ad-hoc* solution. According to Lamport's Bakery algorithm [32], a possible solution is to make both forks available for only a philosopher at a time, by imposing an order of access of the philosophers to the forks. To this end, a new ASM, called *scheduler*, can be added to the DASM so that it decides the order in which the philosophers access the forks. The *scheduler* agent assigns tickets to each philosopher, compare tickets, and let the philosopher with the smallest one to access the forks. All philosophers then access the forks alternately. In order to accomplish this, *scheduler* can set the new monitored function *isMyTurn*: *philosophers* → *boolean*, which states whether a philosopher has the possibility to access the forks or not. Note that the function *isMyTurn* is monitored because only *scheduler* can update it. The model of the single agent can then be refined by changing RULE 1 into RULE 1' as follows:

RULE 1'
**if** *isMyTurn*(**self**) ∧ *owner*(*rightFork*(**self**)) = *undef* ∧ *owner*(*leftFork*(**self**)) = *undef* **then** {
 *owner*(*rightFork*(**self**)) := **self**
 *owner*(*leftFork*(**self**)) := **self**
}

RULE 1' is still vulnerable. The rule can be applied only when the current state is (s.1), which satisfies the `thinking` predicate (still risky). Feature (f.1) holds in that the condition guarding RULE 1' is the logical conjunction of two instances of the shared risky function *owner* and of the monitored function *isMyTurn*; feature (f.2) holds in that the truth value of `thinking` is changed only as a result of the execution of RULE 1' itself. However, the vulnerability of RULE 1' does not imply starvation, because the satisfaction of its condition is guaranteed by the fair behavior of *scheduler*. When *isMyTurn*(**self**) evaluates to *true*, in state (s.1) the locations *owner*(*rightFork*(**self**)) and *owner*(*leftFork*(**self**)) surely evaluate to *undef*, so the computation of the ASM surely evolves towards (s.5).

For the purposes of the present work, it is not necessary to further detail the *scheduler* agent. The interested reader can find the full ASM specification of the Bakery algorithm in [13].

## 6      Evaluation

In order to show the applicability of the proposed approach to a more realistic scenario, we consider the same case study taken into account in [7] and [9], which is here re-modeled in accordance with the newly introduced concepts. This scenario concerns the Ad-hoc On-demand Distance Vector (AODV) routing protocol [35] for Mobile Ad-hoc NETworks (MANETs) [1].

Briefly, MANETs are wireless networks designed for communications among nomadic hosts in absence of fixed physical infrastructure. Hosts can arrange themselves without conforming to a predefined topology; moreover, during their lifetime, they can enter or leave the network at will and continuously change their relative position. Since each host can directly communicate only with its neighbors, MANETs require specific routing protocols capable to take into account the contribution of intermediate hosts to realize communication session between initiator/destination pairs.

Well-known example of routing protocol for MANETs is the AODV protocol. When an initiator node needs a route to a destination node, it starts a so-called *route discovery* process by broadcasting route request (RREQ) packets to all its neighbors. The process is reiterated until the RREQ reaches an intermediate node $n$ which is in the destination's neighborhood or whose routing table contains a fresh route to destination, where the "freshness" is expressed by sequence numbers stored in packets. If so, node $n$ unicasts a route reply (RREP) packet back to initiator and the communication session can start. Otherwise, the route discovery fails when a previously set timeout expires while initiator is waiting for RREPs.

The discussion about the ASM-based model of AODV can be found in [9]. In that work, two formalizations of the protocol were given. The first one is starvation-prone because it intentionally ignores the timeout mechanism adopted to escape infinite waiting: it arises when direct or indirect links between initiator and destination lack. The second one is a refinement which makes the model starvation-free thank to the re-introduction of the timeout. For the purposes of the present work, we here consider the behavior of a node when it acts as initiator, since this is the case that can lead to starvation.

A MANET adopting AODV can be modeled by a DASM including a set of *hosts* = {$h_1$, …, $h_n$}, where each $h_i$ models the behavior of a single node executing the protocol. Each host includes the following functions: *neigbh*: *hosts* → PowerSet(*hosts*), which is a monitored function specifying the neighborhood of a given host; and *wishToInitiate*: *hosts* × *hosts* → *boolean*, which is a monitored function indicating whether a new communication session to a destination *dest* is required for an initiator *init*. Both functions are assumed to be updated by the environment, in accordance with the adopted mobility model and the need to start a communication, respectively. Moreover, to model broadcasting and unicasting of packets, each host is associated with two queues of messages, *requests* and *replies*, which include RREQ and RREP packets, respectively. Finally, each host is associated with a *routing table*: it stores information about the already known routes.

The ASM parameterized rule expressing the behavior of an initiator node is shown in the following. Note that it includes the local function *waiting*: *hosts* × *hosts* → *boolean*, which is a controlled function acting as a flag indicating if initiator is still waiting for (at least) an RREP directed to it. This function allows us to define the

`waiting` predicate over the states, expressing the waiting for RREPs. This predicate is characterized by the value *true* for *waiting*(self, dest).

*Initiator*(*dest*) =
    **RULE 1**: **if** *dest* ∈ *neighb*(**self**) ∨ *dest* ∈ *routingTable*(**self**) **then** {
        *CommunicationSession*(*dest*)
        *wishToInitiate*(**self**, *dest*) := *false*
    }
    **RULE 2**: **if** *dest* ∉ *neighb*(**self**) ∧ *dest* ∉ *routingTable*(**self**) **then** {
        create *RREQ*
        *BroadcastRREQ*(*RREQ*)
        *waiting*(**self**, *dest*) := *true*
    }
    **RULE 3**: **if** *waiting*(**self**, *dest*) **then** {
        **if** *RouteFound*(**self**, *replies*(**self**)) **then** {
            *RREP* = choose *r* ∈ *replies*(**self**) with max sequence number
            *UpdateRoutingTable*(**self**, *RREP*)
            *CommunicationSession*(*dest*)
            *wishToInitiate*(**self**, *dest*) := *false*
            *waiting*(**self**, *dest*) := *false*
            remove RREPs to **self** from *replies*(**self**)
        }
    }

In the pseudo-code above, *RouteFound* returns *true* if an RREP directed to initiator is found in its *replies* queue.

If a route to *dest* is known (i.e. the condition of RULE 1 is fulfilled), the communication session simply starts; otherwise, the parameterized rule *BroadcastRREQ* is executed (RULE 2). Its result consists in inserting a new RREQ into the *requests* queue of all neighbors. When an RREP directed to initiator is received (i.e. *RouteFound*(**self**, *replies*(**self**)) evaluates to *true*), then the computation continues: the communication session starts and the *replies* queue is emptied; otherwise nothing happens, i.e. the node doesn't take any action regarding this particular route discovery attempt, but simply waits.

RULE 3 in the model above states that the agent must wait until it does not receive an RREP corresponding to the RREQ previously sent. This is the vulnerable rule of the algorithm which can make the agent starve, because its computation cyclically returns to states satisfying the `waiting` predicate over the states. In fact, in all other cases, the agent behaves normally, i.e. it processes all control packets coming in and is receptive for requests for other destinations.

If no other ASM sends an RREP back to the agent, *RouteFound*(**self**, *replies*(**self**)) never evaluate to *true*, so the agent's computation cannot evolve. The `waiting` predicate is risky, being defined over the risky location *waiting*(**self**, *dest*). This location, in turn, is risky because is a controlled function determined by the values assumed by the shared queue *replies*. RULE 3, therefore, is vulnerable because its condition includes the risky function *waiting* and the truth value of `waiting` changes only as a result of the execution of RULE 3 itself.

The starvation issue can be solved by a refinement in accordance with the original formulation of the protocol [35], where the authors use a timeout-based mechanism to escape infinite waiting in case a route does not exist. From the ASM point of view, the modification affects only the *Initiator* parameterized rule shown above. More precisely, we need to add the local controlled function *timeout*: *hosts* × *hosts* → *integer*, which models the waiting time for RREPs. Moreover, the change in the machine consists in modifying RULE 3 as follows and in adding the new RULE 4:

RULE 3'
    **if** *waiting*(**self**, *dest*) **then** {
        **if** *RouteFound*(**self**, *replies*(**self**)) **then** {
            *RREP* = choose *r* ∈ *replies*(**self**) with max sequence number
            *UpdateRoutingTable*(**self**, *RREP*)
            *CommunicationSession*(*dest*)
            *wishToInitiate*(**self**, *dest*) := *false*
            *waiting*(**self**, *dest*) := *false*

<div style="margin-left:2em;">

**RULE 4**
```
                    remove RREPs to self from replies(self)
                }
                timeout(self, dest) := timeout(self, dest) – 1
            }
            if waiting(self, dest) ∧ timeout(self, dest) = 0 then {
                wishToInitiate(self, dest) := false
                waiting(self, dest) := false
            }
```

</div>

In order to verify the absence of starvation, we verify that RULE 3' is not vulnerable. To this end, let's note that the predicate `waiting` is no longer risky, because it is defined over the function *waiting*. In turn, this function is no longer risky because, thanks to RULE 4, its value is determined by the new controlled function *timeout*. Finally, note that *timeout* is not a risky function because it is controlled and its value is not determined by any other function. In fact, its value is set, within a finite amount of time, only by the ASM it belongs to: under the assumption that the initial value of *timeout* is greater than 0, it decreases at each computational step, so surely converging to 0.

## 7    Conclusion

Abstract State Machines are very helpful for modeling critical and complex systems; however, unlike other well-known formalisms, they lack of inherent characterizations of properties interesting from a computational point of view, in particular starvation-free. The aim of this paper was to reinforce the ASM framework as a conceptual tool for studying starvation. Firstly, starting from some acknowledged informal definitions of starvation, we have stated a general necessary condition. Then, this condition has been translated in terms of ASMs, so enriching the ASM framework with the capability of capturing starvation issues in an operational fashion. To this end, we have capitalized and refined two works in which the concepts of *predicate abstraction over ASM states* [8] and *vulnerable rule* [9] were proposed as tools capable of capturing the risk of starvation inside ASMs. More precisely, we have proved that the presence of vulnerable rules is a necessary condition for starvation.

Thanks to these notions, developers can recognize the risk of starvation in an ASM model of a system, so they can re-model it in advance, before its development, with evident effort savings.

Nevertheless, our approach also presents drawbacks. Because of decidability issues, it cannot be completely automatized. Furthermore, as other manual techniques, it is human-based, so is error-prone and requires expertise in order to find an appropriate abstraction of the system to be verified. In other words, any analysis is as good as the model is.

The general discussion has then been completed by the application to the well-known Dining Philosophers problem as running example. It is worth remarking that, although simple, this problem is adequately general in that it enables an abstract description of starvation issues that can easily be applied to any specific domain. Furthermore, in order to show the applicability of the proposed approach to a real case, the starvation issue has been investigated in a scenario taken from the MANET domain.

The ASM-characterization of starvation provided in this paper refers to the analysis of dependences within and between state machines, but this issue is only statically expressed in terms of rules, and is not deepened. Future directions of the research should investigate such dependencies, taking advantages, for example, of some existing tools. Two remarkable references are [3], where a tool based on Extended Finite State Machines is used to study several forms of dependence, included non-termination; and by [15], where Finite State Machines form the basis of a verification tool capable of verifying interaction-oriented properties of interactive software components.

Since, in the present paper, we have focused on the starvation analysis, it represents a step in our research aimed at characterizing computationally interesting properties within the ASM framework. Future work will move from starvation-free to deadlock-free, according to Tanenbaum's remark about the distinction between starvation and deadlock [44]. A deadlock happens when there is a circular chain in the resources request, so that all processes are mutually blocked and the entire system is unable to make progress [41]. Some results on applying ASMs to deadlock-freedom analysis appeared in [10].